\documentclass[fleqn,usenatbib]{mnras}
\pdfoutput=1
\usepackage{mathtools}
\usepackage{appendix}
\usepackage{bm}
\usepackage{soul}
\usepackage{natbib}
\usepackage{orcidlink}
\usepackage{newtxtext,newtxmath}
\usepackage[T1]{fontenc}
\usepackage{ae,aecompl}

\defcitealias{Tannock_2021AJ}{T21}
\defcitealias{Lipatov_2020ApJ}{LB20}

\title{Effects of Rotation on the Spectra of Brown Dwarfs}

\author[Lipatov, Brandt, \& Batalha]{
Mikhail Lipatov,$^{1}$\thanks{E-mail: mikhail@physics.ucsb.edu}\orcidlink{0000-0001-9939-1758}
Timothy D.~Brandt,$^{1}$\orcidlink{0000-0003-2630-8073}
and Natasha E.~Batalha$^{2}$\orcidlink{0000-0003-1240-6844} \\
$^{1}$Department of Physics, University of California, Santa Barbara, CA 93106, USA \\
$^{2}$NASA Ames Research Center, Mountain View, CA 94043, USA
}

\pubyear{2022}

\begin{document}
\label{firstpage}
\pagerange{\pageref{firstpage}--\pageref{lastpage}}
\maketitle

\begin{abstract}
Measured rotational speeds of giant planets and brown dwarfs frequently constitute appreciable fractions of the breakup limit, resulting in centrifugal expansion of these objects at the equator. According to models of internal energy transport, this expansion ought to make the poles of a rotator significantly hotter than the equator, so that inclination of the rotational axis greatly affects both spectral shape and total flux. In this article, we explore the dependence of a substellar object's observables on its rotational speed and axis inclination. To do so, we combine \texttt{PICASO} (a Planetary Intensity Code for Atmospheric Spectroscopy Observations) with software \texttt{PARS} (Paint the Atmospheres of Rotating Stars). The former computer program models radiative transfer within plane-parallel planetary atmospheres, while the latter computes disk-integrated spectra of centrifugally deformed gaseous masses. We find that the specific flux of a typical fast-rotating brown dwarf can increase by as much as a factor of 1.5 with movement from an equator-on to a pole-on view. On the other hand, the distinctive effect of rotation on spectral shape increases toward the equator-on view. The latter effect also increases with lower effective temperature. The bolometric luminosity estimate for a typical fast rotator at extreme inclinations has to be adjusted by as much as $\sim20\%$ due to the anisotropy of the object's observed flux. We provide a general formula for the calculation of the corresponding adjustment factor in terms of rotational speed and inclination.
\end{abstract}

\section{Introduction} \label{introduction}

\subsection{Brown Dwarf Structure and Evolution}

When a portion of a giant molecular cloud collapses to form a single gaseous object, the nature and the fate of that object are mainly determined by its mass. Above $75\,M_{\rm Jup}$, the object is massive enough that its core temperature and pressure enable the fusion of hydrogen, ${\rm ^1 H}$ \citep{Basri_2000_ARA&A}. This, by definition, makes the object a main-sequence star, whose temperature and luminosity remain relatively constant for billions of years -- except in the rare case that the star's mass significantly exceeds $1\,M_\odot$. Between $13\,M_{\rm Jup}$ and $75\,M_{\rm Jup}$, however, the collapsed object cannot fuse ${\rm ^1 H}$, although it can still fuse deuterium, ${\rm ^2 H}$ \citep{Saumon_1996ApJ}. This sort of an object is a brown dwarf (BD). Infra-red excess from young brown dwarfs indicates the presence of accretion disks. This confirms that BDs are formed via the above-mentioned process of molecular cloud collapse, much like stars \citep{Muench_2001ApJ,Natta_2001A&A}. Regardless of their formation mechanism, objects with mass below $13\,M_{\rm Jup}$ never achieve nuclear fusion in the core and are classified as planets.

The objects that are intermediate between stars and planets in terms of mass -- brown dwarfs -- are particularly interesting. No such object exists in our own solar system, though the latter does contain several giant planets and a star. 
According to the theory of brown dwarf formation and evolution, the surface effective temperature of a typical newly formed BD is in the neighborhood of $\sim$2,$500 - 3$,$000\,{\rm K}$ \citep{Chabrier_2000ARA&A}. However, this temperature drops relatively quickly, reaching about 50\% of its original value by $\sim$500\,Myr as the deuterium content dwindles, even for massive brown dwarfs. The cooling process leads to observed brown dwarf temperatures that are as low as $\sim$250\,K for low masses and old ages \citep{Luhman_2014ApJ,Phillips_2020A&A}.

If we approximate brown dwarfs as black bodies, Wien's displacement law tells us that these objects' surface temperature drop corresponds to a rise in peak spectral wavelength from ${\rm \sim1\,\mu m}$ well into the infra-red portion of the spectrum. In addition, the temperature drop corresponds to a drop in luminosity. In particular, luminosities of the heaviest BDs decrease past the luminosities of the least massive main-sequence stars around ${\rm 1\,Gyr}$ \citep{DAntona_1985ApJ}. It wasn't until relatively recently that astronomical detection technology enabled the observation of such low luminosities in the infra-red. Accordingly, although the existence of brown dwarfs was predicted in the early 1960's \citep{Kumar_1963ApJ,Hayashi_1963PThPh} and although BDs are about as common as stars \citep{Bate_2002MNRAS}, the first object unambiguously classified as a brown dwarf was announced as late as 1995 \citep{Basri_2000_ARA&A,Nakajima_1995Nature,Oppenheimer_1995Science}.

Models of brown dwarf structure and evolution predict that the radii of these objects are generally close to Jupiter's and that these radii decrease with object age as their interiors cool and electron degeneracy becomes an increasingly important source of pressure support \citep{Baraffe_2003A&A,Saumon_2008ApJ}. The former prediction agrees with the observed brown dwarf radii, which are anywhere between $\sim0.6\,M_{\rm Jup}$ and $\sim1.4\,M_{\rm Jup}$ \citep{Sorahana_2013ApJ,Carmichael_2020AJ}.

Due to their relatively low temperatures, brown dwarf atmospheres exhibit a number of complexities that are not present in the atmospheres of stars. In particular, BD spectra deviate strongly from black body spectra due to prominent molecular absorption features \citep{Allard_1997ARA&A}. Furthermore, clouds can form in the potentially rotating atmospheres, adding further complexity \citep{Ackerman_2001ApJ,Lew_2020AJ,Tan_2021MNRAS}. At any given time point, such atmospheric properties strongly depend on surface effective temperature. The atmosphere of a brown dwarf, in turn, determines the shape of its spectrum and regulates its cooling \citep{Chabrier_2000ARA&A,Allard_1997ARA&A}.

\subsection{Rotating Brown Dwarfs}

Brown dwarfs spin at a significant rate due to the angular momentum that they inherit from their progenitor molecular clouds. Specifically, observed rotation periods of newly born BDs are $\sim100\,{\rm hr}$, falling to $\sim10\,{\rm hr}$ by $\sim100\,{\rm Myr}$ due the above-mentioned cooling and contraction process \citep{Eisloffel_2007IAUS,Crossfield_2014A&A}. The rotational periods of the fastest-rotating brown dwarfs are as low as $\sim 1\,{\rm hr}$, corresponding to significant fractions (up to a third) of the Keplerian breakup velocity \citep{Tannock_2021AJ}. Rotation rates at appreciable fractions of the breakup limit cause centrifugal deformation and surface temperature variation in stars \citep{Monnier_2007_science,Desouza_2014_A&A}. This, in turn, causes the observed spectra, magnitudes, and bolometric fluxes of stars to depend on the inclination of the rotational axis with respect to the observer \citep[][hereafter LB20]{Lipatov_2020ApJ}. Specifically, the intensity of a typical fast rotating star varies by as much as a factor of 1.5 between the two extreme inclinations in spectral regions with appreciable flux, which corresponds to $\gtrsim0.3$ in visual magnitude difference. Differences in color and spectral line shape are potentially detectable as well.  Rotation is expected to be similarly appreciable in brown dwarfs \citep[e.g.,][]{Sanghavi_2018ApJ}. 

Inference of brown dwarf parameters based on model atmospheres yields effective temperatures, surface gravities, radii, and metallicities of these objects \citep{Zhang_2021ApJ}. Currently, such inference is based on nonrotating, spherically symmetric models with uniform surface temperature, even for objects known to be rotating at significant fractions of the breakup limit \citep{Tannock_2021AJ,Eisloffel_2007IAUS,Crossfield_2014A&A}. This can lead to inaccurate estimation of temperatures and luminosities from BD spectra and bolometric fluxes, since the latter observables depend on rotational axis inclination and since the temperature of a rotating brown dwarf varies significantly across its surface. In this work, we quantify the dependence of BD observables on rotational speed and object orientation, as well as the associated effect on inferred nonrotating model parameters. To accomplish this, we utilize Sonora brown dwarf 1D climate and chemistry models \citep{Marley2021ApJ}, the \texttt{PICASO} spectroscopy code \citep{batalha2019}, and \texttt{PARS} (Paint the Atmospheres of Rotating Stars) -- software that computes observed spectra of self-gravitating, rotating gaseous objects \citepalias{Lipatov_2020ApJ}.

The rest of this article is organized as follows. Section \ref{distributions} describes our assumptions regarding brown dwarf rotation and internal mass distribution. Section \ref{spectra} details the calculation of the synthetic spectra. Section \ref{bd_models} lists the parameters of the brown dwarf models whose spectra we calculate. Section \ref{rotation_effects} describes quantitative comparison between rotating and nonrotating model spectra, as well as a relationship between bolometric flux and inferred luminosity for a rotating dwarf. Section \ref{results} reports our findings -- the effect of rotation on brown dwarf spectra and inferred luminosities. We discuss these results and conclude the article in Section \ref{discussion}.

\section{Rotation and Mass Distribution Models} \label{distributions}

In this section, we introduce our rotation model and two mass distribution models for brown dwarfs. Of these, the rotation model and the Roche model of mass distribution were previously shown to accurately predict spectra of A-type rotating stars \citepalias{Lipatov_2020ApJ}.

We assume that a given brown dwarf rotates at some uniform angular speed $\Omega$. The resulting centrifugal forces cause the object's equatorial radius to be greater than its polar radius. In other words, rotation leads to non-zero oblateness $f$, a quantity that we define as
\begin{equation} \label{eq:f}
    f\equiv \frac{R_{\rm e} - R_{\rm p}}{ R_{\rm e}},
\end{equation}
where $R_{\rm e}$ is the dwarf's equatorial radius and $R_{\rm p}$ is its polar radius. Re-arrangement of Equation \eqref{eq:f} yields the polar radius as a function of equatorial radius and oblateness:
\begin{equation} \label{eq:Rp}
    R_{\rm p} = R_{\rm e} (1 - f).
\end{equation}

Another useful quantity to consider is the dimensionless rotational velocity, $\omega \equiv \Omega / \Omega_{\rm K} \in [0, 1)$. Here, $\Omega_{\rm K}$ is the Keplerian rotational limit -- the maximum angular velocity that the rotator can have without breaking up. 

In general, there are no simple expressions for either $\omega$ or $\Omega_{\rm K}$ in terms of the dwarf's fundamental parameters such as mass, radius, or oblateness. On the other hand, useful and relatively simple expressions for $\omega$ and $\Omega_{\rm K}$ exist under the assumption of the Roche mass distribution, which places all mass at the brown dwarf's center. Given the Roche distribution, the Keplerian limit is
\begin{equation} \label{eq:OmegaK}
    \Omega_{\rm K} = \frac{G M}{R_{\rm e}^3}.
\end{equation}
The Roche distribution further implies Equation (2) in \citetalias{Lipatov_2020ApJ}, 
\begin{equation} \label{eq:Rrat}
    \frac{R_{\rm e}}{R_{\rm p}} = 1 + \frac{\omega_{\rm R}^2}{2},    
\end{equation}
where $\omega_{\rm R}$ is the theoretical value of $\omega$ that yields the observed ratio of equatorial and polar radii under the assumption of the Roche distribution. Relatedly, $\omega_{\rm R}$ yields the observed oblateness, as well. To see this, we combine Equation \eqref{eq:Rrat} with Equation \eqref{eq:f} to get oblateness in terms of $\omega_{\rm R}$,
\begin{equation} \label{eq:f_R}
    f(\omega_{\rm R}) = \frac{1}{1 + 2/\omega_{\rm R}^2}.
\end{equation}
We then invert Equation \eqref{eq:f_R} to obtain
\begin{equation} \label{eq:omegaR}
    \omega_{\rm R}(f) = \sqrt{\frac{2 f}{1 - f}}.
\end{equation}

Relaxation of the Roche distribution assumption allows for a more realistic internal mass distribution. Under these conditions and a certain set of approximations, the Darwin-Radau relation connects the moment of inertia of a brown dwarf to its oblateness and rotational speed \citep{Barnes_2003ApJ}:
\begin{equation} \label{eq:C}
    \mathbb{C} = \frac{C}{MR_{\rm e}^2} = \frac{2}{3} \left[1-\frac{2}{5}\left(\frac{5}{2} \frac{\omega^2}{f} - 1\right)^{1/2}\right].
\end{equation}
Here, $C$ is the moment of inertia about the rotational axis and $\mathbb{C}$ is the corresponding dimensionless quantity. Spacecraft gravity field observations estimate that $\mathbb{C} = 0.276$ for Jupiter \citep{Ni_2018AA}. The Darwin-Radau relation, in combination with the observed oblateness and rotational velocity, yields $\mathbb{C} = 0.22$ for Saturn \citep{Fortney_2018book}. For a uniform-density sphere, $\mathbb{C} = 0.4$. We can re-write Equation \eqref{eq:C} to obtain $f$ in terms of $\omega$ and $\mathbb{C}$:
\begin{equation} \label{eq:f_DR}
    f(\omega, \mathbb{C}) = \frac{40\, \omega^2}{116 - 300\, \mathbb{C} + 225\, \mathbb{C}^2}.
\end{equation}

When $\mathbb{C} = 0$, Equation \eqref{eq:f_DR} reduces to $f = \omega^2 / 2.9$. For $\omega \in [0, 1)$ and $\omega \equiv \omega_{\rm R}$, this equation is numerically similar, although not identical, to Equation \eqref{eq:f_R}. Although the latter assumes a mass distribution with $\mathbb{C} = 0$, it does so under a different set of approximations. 

Finally, we can re-write Equation \eqref{eq:C} to express dimensionless rotational speed in terms of oblateness and the moment of inertia:
\begin{equation} \label{eq:omega_DR}
    \omega(f, \mathbb{C}) = \sqrt{\frac{1}{40}\,f\,(116 - 300\,\mathbb{C} + 225\,\mathbb{C}^2)}.
\end{equation}

In the rest of this article, we denote the observed rotational speed of brown dwarfs by $\omega$ and assume that it relates to observed oblateness via the Darwin-Radau relation. On the other hand, at times we will also calculate $\omega_{\rm R}$ -- the rotational speed that yields the observed oblateness under the assumption of the Roche model.

\section {Computation of Spectra} \label{spectra}

\subsection{Spectral Intensity as a Function of Surface Parameters}

We compute plane-parallel-derived spectra using the  pressure-temperature, and abundance profiles from the Sonora cloud-free solar metallicity grid \citep{Marley2021ApJ}, which is available on Zenodo \citep{marley_mark_2018}. The cloud-free atmospheres exist on a grid of gravity and effective temperature in the range of: 3.25$\le$logg$\le$5.5 and 200$\le$T$_\mathrm{eff}\le$2400~K. Therefore, they are particularly relevant for relatively low temperatures, such as those associated with Model 880 in Section \ref{variants}, since we do not expect clouds at these temperatures. Even though higher brown dwarf temperatures are associated with clouds, we use the cloud-free atmospheres for higher-temperature models as well, since our main goal is to show the qualitative effect of rotation on brown dwarf spectra.

Using the Sonora pressure-temperature, and abundance profiles, we compute the specific intensity in ${\rm erg\,s^{-1}\,Hz^{-1}\,sr^{-1}\,cm^{-2}}$ on a grid of the cosine of the viewing angle $\mu = \cos \phi$, wavelength $\lambda$, surface effective temperature, and surface gravitational acceleration using version 2.3 of the open source code \texttt{PICASO} \citep{batalha2019,picaso2022}, which has previously been used to compute the thermal emission spectra of brown dwarfs \citep[e.g.][]{Mang2022ApJ} and exoplanets \citep[e.g.][]{Robbins2022}.

\texttt{PICASO} uses the same radiative transfer methodology \citep{Marley1999} and opacities \citep{Freedman2014ApJS,lupu_roxana_2022_6600976} as those used to compute the final pressure-temperature profiles, described in \citet{Marley2021ApJ}. Specifically, it uses the \citet{Toon1989JGR} source function technique and includes the molecular and atomic opacities for ${\rm C_2 H_2}$, ${\rm C_2 H_4}$, ${\rm C_2 H_6}$, ${\rm CH_4}$, ${\rm CO}$, ${\rm CO_2}$, ${\rm CrH}$, ${\rm Cs}$, ${\rm FeH}$, ${\rm H_2 O}$, ${\rm H_2 S}$, ${\rm HCN}$, ${\rm K}$, ${\rm Li}$, ${\rm LiCl}$, ${\rm MgH}$, ${\rm N_2}$, ${\rm NH_3}$, ${\rm Na}$, ${\rm OCS}$, ${\rm PH_3}$, ${\rm Rb}$, ${\rm SiO}$, ${\rm TiO}$, and ${\rm VO}$. They also include continuum opacities for the following interactions: ${\rm H_2-H_2}$, ${\rm H_2-He}$, ${\rm H_2-H}$, ${\rm H-}$, ${\rm H_2-}$.

For each gravity-effective temperature pair available in the Sonora grid, we compute spectra for viewing angles, $\mu$, of: [0.09, 0.18, 0.28, 0.37, 0.45, 0.54, 0.61, 0.68, 0.75, 0.81, 0.86, 0.9, 0.94, 0.97, 0.98, 0.99]. The $\lambda$ grid consists of 9831 wavelengths between 0.7 and 5 ${\rm \mu m}$, with spectral resolution $R = 5000$. This wavelength grid becomes the grid of our brown dwarf spectra. The temperature range, which is equivalent to that of the Sonora grid, is between 200 and 2400 K, with temperature resolution equal to 25 K below 600 K, 50 K between 600 and 1000 K, and 100 K above 1000 K. The gravity grid is [17, 31, 56, 100, 178, 316, 562, 1000, 1780, 3160] ${\rm m\,s^{-2}}$.

We use $\mu$ equal to 0.1 and 0.4 as the boundaries of the viewing angle intervals for the piecewise interpolation of intensity as a function of $\mu$ by \texttt{PARS} \citepalias{Lipatov_2020ApJ}. These viewing angle intervals proved to be optimal in the application of \texttt{PARS} to stars.

\subsection{Disk-Integrated Spectra}

To compute disk integrated spectra of rotating brown dwarfs, we pass the plane-parallel-derived spectra to Paint the Atmospheres of Rotating Stars (\texttt{PARS}) -- software that quickly and accurately computes spectra of self-gravitating rotating gaseous objects such as stars and brown dwarfs \citepalias{Lipatov_2020ApJ}. \texttt{PARS} computes the surface shape and the local effective temperature everywhere on the surface under the assumption of the Roche mass distribution, then integrates specific intensities over the surface to get a spectrum. 

One of the inputs to \texttt{PARS} is the dimensionless rotational velocity we discuss in Section \ref{distributions}. The gravity darkening effects we wish to compute are largely a function of geometric distortion, epitomized by oblateness. Thus, given a model with oblateness $f$, we calculate velocity $\omega_{\rm R}$ using Equation \eqref{eq:omegaR} and provide \texttt{PARS} with this velocity value. This ensures that the \texttt{PARS} model has the correct oblateness. Other brown dwarf parameters that serve as part of the input to \texttt{PARS} are mass $M$, equatorial radius $R_{\rm e}$, luminosity $L$ and distance. For all our models, we adopt a fiducial distance of 10 parsecs. Table \ref{tab:bd_params} and Section \ref{bd_models} describe the process that yields the remaining parameters.

\section{Brown Dwarf and Exoplanet Case Studies} \label{bd_models}

We construct model brown dwarfs that span a range of rotation rates and effective temperatures; we ultimately compute their synthetic spectra as described in Section \ref{spectra}.  Our models are inspired, in particular, by two very fast-rotating
brown dwarfs in the literature, J0348-6022 and J0407+1546 \citep[][hereafter T21]{Tannock_2021AJ}.  
We also construct a model inspired by $\beta$ Pictoris b, one of the first imaged exoplanets \citep{Lagrange_2009A&A,Lagrange_2010Sci} and an object near the deuterium-burning boundary
\citep{Chilcote_2017AJ,Brandt_2021AJ}. The upper half of Table \ref{tab:bd_params}, above the mid-table horizontal line, lists the literature-based parameters of the three above-mentioned substellar objects. The lower half lists the parameters we assume or derive. Bold text indicates the parameters that serve as input to the synthetic spectrum software, \texttt{PARS}. In this section, we examine the process that yields all the parameters in Table \ref{tab:bd_params}. 

\begin{table*}
\centering
\caption[]{Parameters of our three main substellar models. }
\begin{tabular}{lccc}
\hline
    Name & 
    J0348-6022 (Model 880) & 
    J0407+1546 & 
    $\beta$ Pictoris b \\[2pt]
    \hline
Parameters Reference & \cite{Tannock_2021AJ} & \cite{Tannock_2021AJ} & \cite{Chilcote_2017AJ} \\[2pt]
Spectral Type & T7 & L3.5 & L2 \\[2pt]
$\mathbf{M\,(M_{\rm Jup})}$ & \textbf{42.9} & \textbf{67.0} & \textbf{12.9} \\[2pt]
$\mathbf{R_{\rm e}\,(R_{\rm Jup})}$ & \textbf{0.905} & \textbf{0.973} & \textbf{1.46} \\[2pt]
$\bar{T}_{\rm eff}\,({\rm K})$ & 880 & 1840 & 1724 \\[2pt]
$P_{\rm rot}\,({\rm h})$ & 1.08 & 1.23 & \\[2pt]
$v \equiv v_{\rm e}\sin{i}\,{\rm (km/s)}$ & & & 25\\
$i$ & & &${\pi/2}$~\textsf{$^{a}$}\\[2pt]
$f \equiv \frac{R_{\rm e} - R_{\rm p}}{R_{\rm e}}$ & 0.08 & 0.05 & 0.029~\textsf{$^{b}$} \\[2pt]
\hline
$R_{\rm avg} \equiv \frac{R_{\rm e} + R_{\rm p}}{2} ~(R_{\rm Jup})$ & 0.869 & 0.949 & 1.44\\[2pt]
${\bf L} = 4 \pi R_{\rm avg}^2 \sigma_{\rm SB} \bar{T}_{\rm eff}^4 ~ {\bf (L_\odot)}$ & ${\bf 4.3\times10^{-6}}$ & ${\bf 9.8\times10^{-5}}$ & ${\bf 1.7\times10^{-4}}$\\[2pt]
$\Omega = 2\pi/P_{\rm rot}~({\rm rad\,s^{-1}})$ & $0.0016$ & $0.0014$ & \\[2pt]
$\Omega = v / (R_{\rm e}\,\sin{i})~({\rm rad\,s^{-1}})$ & & & $0.00024$\\[2pt]
$\Omega_{\rm K} = \sqrt{GM/R_{\rm e}^3}~({\rm rad\,s^{-1}})$ & $0.0045$ & $0.0050$ & $0.0012$\\[2pt]
$\omega = \Omega / \Omega_{\rm K}$ & $0.36$ & $0.28$ & $0.20$\\[2pt]
$\mathbb{C} = C/(MR_{\rm e}^2) $ & $0.20$~\textsf{$^{c}$} & & $0.25$~\textsf{$^{d}$}\\[2pt]
${\bf \boldsymbol\omega_{\rm R} = 2\,f / (1 - f)}$~\textsf{$^{e}$} & ${\bf 0.42}$ & ${\bf 0.32}$ & ${\bf 0.24}$ \\[2pt] \hline
\end{tabular}
\begin{list}{}{}
\item[\textsf{$a$}] Assuming that the direction of the brown dwarf's spin angular momentum is the same as those of its star's spin, orbital, and circumstellar disk angular momenta \citep[e.g.,][]{Kraus_2020ApJ}.
\item[\textsf{$b$}] From $\omega$ and the Darwin-Radau relation, assuming $\mathbb{C}=0.25$.
\item[\textsf{$c$}]Computed from $f$ and $\omega$ according to Equation \eqref{eq:C}.
\item[\textsf{$d$}] Similar to estimates for Jupiter \citep{Ni_2018AA} and Saturn \citep{Fortney_2018book}.
\item[\textsf{$e$}] Dimensionless rotational velocity $\omega$ corresponding to oblateness $f$ under the Roche model.
\end{list}
\label{tab:bd_params}
\end{table*}

\subsection{J0348-6022}

Object 2MASS J03480772-6022270, a.k.a.~J0348-6022, is a rapidly rotating brown dwarf with spectral type T7. Table \ref{tab:bd_params} presents \citetalias{Tannock_2021AJ}'s estimates of this dwarf's mass $M$, equatorial radius $R_{\rm e}$, rotational period $P_{\rm rot}$, oblateness $f$, and surface temperature $\bar{T}_{\rm eff}$. We assume the surface temperature estimate to be some average over the gravity-darkened dwarf surface. Our work does not aim to draw specific conclusions about this or other astrophysical objects. Instead, our goal is to demonstrate the effects of rotation on the spectra of such objects in general. Thus, our conclusions do not require object parameter values that are more exact than the ones we present. All our brown dwarf models inspired by J0348-6022 have the same mass, $M = 0.041\,M_\odot$.

Equation \eqref{eq:Rp} yields the polar radius $R_{\rm p}$ of our J0348-6022 model. With $R_{\rm avg} = (R_{\rm e} + R_{\rm p}) / 2$, we set the dwarf's luminosity $L$ to
\begin{equation} \label{eq:lum_avg}
    L = 4 \pi R_{\rm avg}^2 \sigma_{\rm SB} \bar{T}_{\rm eff}^4,
\end{equation}
where $\sigma_{\rm SB}$ is the Stefan-Boltzmann constant.  In Equation \eqref{eq:lum_avg}, $4 \pi R_{\rm avg}^2$ is close to, but not equal to, the surface area of the star; $\bar{T}_{\rm eff}$ represents a characteristic effective temperature.

We also compute the rotation speed as $\Omega = 2\,\pi / P_{\rm rot}$, the Keplerian limit $\Omega_{\rm K}$ via Equation \eqref{eq:OmegaK}, the dimensionless rotational speed as $\omega = \Omega / \Omega_{\rm K}$, and the dimensionless moment of inertia from oblateness $f$ and $\omega$ according to Equation \eqref{eq:C}. Furthermore, we use Equation \eqref{eq:omegaR} to compute the rotational speed $\omega_{\rm R}$ that would produce the observed oblateness $f$ under the assumption of the Roche distribution.

\subsubsection{Model Variants} \label{variants}

In order to explore the effects of rotation on the inference of brown dwarf parameters from nonrotating models, we create a set of models based on J0348-6022 with a variety of luminosities and rotational speeds. 

We first consider the J0348-6022 model from Table \ref{tab:bd_params}, with its oblateness $f = 0.08$, dimensionless rotational speed $\omega = 0.36$, dimensionless moment of inertia $\mathbb{C} = 0.20$, Roche model equivalent speed $\omega_{\rm R} = 0.42$, average effective temperature $\bar{T}_{\rm eff} = 880\,{\rm K}$, luminosity $L = 4.3\times10^{-6}\,L_\odot$, and equatorial radius $R_{\rm e} = 0.905\,R_{\rm Jup}$. We call this Model 880 and use \texttt{PARS} software \citepalias{Lipatov_2020ApJ} to determine the effective temperatures $T_{\rm eff}$ of its equator and poles; Section \ref{spectra} describes the function of \texttt{PARS} in more detail. We then determine the luminosities of nonrotating dwarfs that correspond to the equatorial and polar effective temperatures via
\begin{equation} \label{eq:lum}
    L = 4 \pi R_{\rm avg}^2 \sigma_{\rm SB} T_{\rm eff}^4,
\end{equation}
where $R_{\rm avg}$ remains constant, corresponding to constant oblateness and equatorial radius. Finally, we create a set of nonrotating models with effective temperatures (and hence luminosities) that are intermediate between the equatorial and polar values of Model 880. In Section \ref{rotation_effects} we will compare the properties of these nonrotating models to the inclination-dependent observables of the rotating version of Model 880.

We also create a set of models by reducing the rotation rate $\omega_{\rm R}$ of Model 880 more gradually. As $\omega_{\rm R}$ decreases, oblateness decreases according to Equation \eqref{eq:f_R} and so does the difference between the equatorial and polar radii. We keep luminosity and $\bar{T}_{\rm eff}$ constant in Equation \eqref{eq:lum_avg} by holding $R_{\rm avg} = (R_{\rm e} + R_{\rm p}) / 2$ constant as oblateness varies. To achieve this, we substitute for the polar radius according to Equation \eqref{eq:Rrat} in the definition of $R_{\rm avg}$ and obtain 
\begin{equation}
    R_{\rm avg} = \frac{R_{\rm e} + R_{\rm p}}{2} = \frac{R_{\rm e}}{2}\, \frac{2+\omega_{\rm R}^2/2}{1+\omega_{\rm R}^2/2}, 
\end{equation}
so that 
\begin{equation} \label{eq:Re_Ravg}
    R_{\rm e} = 2\,R_{\rm avg}\,\frac{2+\omega_{\rm R}^2}{4+\omega_{\rm R}^2}. 
\end{equation}
Thus, as we reduce $\omega_R$ in discrete steps, we keep $R_{\rm avg}$ constant at the values of Model 880 by setting the equatorial radius according to Equation \eqref{eq:Re_Ravg}. In addition to keeping $\bar{T}_{\rm eff}$ and $L$ constant, we keep $\mathbb{C}$ constant, as well. The new models have oblatenesses given by Equation \eqref{eq:f_R} and actual dimensionless rotational speeds $\omega$ given by Equation \eqref{eq:omega_DR}. 

We next create Model 600, which has $\bar{T}_{\rm eff} = 600\,{\rm K}$, while retaining the geometry, mass distribution, and rotational speed properties of Model 880. Given that Model 600 has $R_{\rm avg} = 0.0893\,R_\odot$, Equation \eqref{eq:lum_avg} tells us that this model's luminosity is $9.3\times10^{-7}\,L_\odot$. We then create nonrotating models at different effective temperatures and rotating models at different rotational speeds that correspond to Model 600 and its average effective temperature, the same way we created such variations for Model 880.

Similarly, we create Models 400 and 1500, which have the effective temperatures implied by their names and respective luminosities equal to $1.8\times10^{-7}\,L_\odot$ and $3.6\times10^{-5}\,L_\odot$, with other parameters the same as in Models 880 and 600. We make nonrotating and rotating model sets that correspond to each new luminosity, as well.

\subsection{Additional Case Studies}

Object 2MASS J04070752+1546457, a.k.a.~J0407+1546, is a rapidly rotating brown dwarf with spectral type L3.5 \citepalias{Tannock_2021AJ}. It serves as a template for a brown dwarf model with parameters that we obtain in the same way we obtain the parameters for Model 880.

Object $\beta$ Pictoris ($\beta$ Pic) b, which is at the giant planet / brown dwarf mass boundary, serves as another model template. We obtain its mass $M$, equatorial radius $R_{\rm e}$, and average temperature $\bar{T}_{\rm eff}$ from \citet{Chilcote_2017AJ}. \cite{Snellen_2014Nature} measure this object's projected rotational velocity $v \equiv v_{\rm e}\,\sin{i}$. \citet{Kraus_2020ApJ} find that the angular momentum vector of the $\beta$ Pic stellar photosphere, the angular momentum of the $\beta$ Pic b orbital movement, and the angular momentum of the system's outer debris disk are well-aligned with mutual inclinations $\le 3 \pm 5^{\circ}$, which indicates that $\beta$ Pic b formed in a system without significant primordial misalignments. Given these findings and the orbit's near edge-on orientation, we adopt an inclination of $i = \pi / 2$ for the planet's spin. Here and in the rest of this article, we define $i$ so that $i = \pi / 2$ corresponds to an equator-on view and $i = 0$ -- to a pole-on view. 

We then obtain the object's angular velocity as $\Omega = v / (R_{\rm e}\,\sin{i})$. At this point, we compute the average radius $R_{\rm avg}$, luminosity $L$, Keplerian limit $\Omega_{\rm K}$, and dimensionless rotational velocity $\omega$ the same way we calculate these parameters for Model 880. We adopt a dimensionless moment of inertia equal to $\mathbb{C} = 0.25$, which is similar to the estimates of this quantity for the giant planets of the Solar system \citep{Ni_2018AA,Fortney_2018book}. Given $\omega$ and $\mathbb{C}$, we use Equation \eqref{eq:f_DR} to calculate an estimate for oblateness $f$. This, in turn, yields the Roche model equivalent velocity $\omega_{\rm R}$ via Equation \eqref{eq:omegaR}.

We list all the parameters for the models based on J0407+1546 and $\beta$ Pictoris b in Table \ref{tab:bd_params}.

\section {Effects of Rotation} \label{rotation_effects}

\subsection{Comparisons to Spectra of Nonrotating Objects} \label{comparisons}

In this section, we address the detectability of gravity darkening from spectra alone.  The spectrum of a rotating brown dwarf will contain contributions from different effective temperatures; it will differ from the spectrum of a spherical brown dwarf no matter the temperature of the latter.  We aim to quantify the difference between the spectrum of a rotating brown dwarf and the best-matching nonrotating spectrum.  This serves as a metric of rotation's spectral detectability that is independent of projected rotational speed $v \sin i$ as measured from line broadening.

In the comparison of a rotating dwarf's flux $x(\lambda)$ to the flux of a nonrotating model $y(\lambda)$, we think of the former as the independent variable and of the latter as the dependent variable. We model $y(x)$ via linear regression. If $\tilde{x} = x - \bar{x}$ and $\tilde{y} = y - \bar{y}$ are the mean-subtracted versions of the two spectra, then the slope of the best-fitting linear model is $b = \sum{\tilde{x}\tilde{y}} / \sum{\tilde{x}^2}$. The root mean squared deviation (RMSD) of this best-fitting line from the rotating model spectrum is ${\rm RMSD}(\hat{\cal F})=\sqrt{\sum{(\tilde{y} - b\tilde{x})^2 / n}}$, where $n$ is the number of wavelengths at which each spectrum is sampled. The units of this deviation are the same as the units of flux. 

\texttt{PARS} produces flux density ${\cal F}(\lambda)$ in ${\rm erg\,s^{-1}\,Hz^{-1}\,cm^{-2}}$ on a grid of wavelengths $\lambda$. We calculate a grid of $\tilde{\nu} = 1/\lambda$, which is proportional to frequency in ${\rm Hz}$. We then integrate ${\cal F}(\tilde{\nu})$ on the grid of $\tilde{\nu}$ using the variable-interval trapezoidal rule. This yields the integrated flux ${\cal F}$ in units that are proportional to ${\rm erg\,s^{-1}\,cm^{-2}}$. We divide ${\cal F}$ by the range of $\tilde{\nu}$ to obtain the average flux $\bar{\cal F}$ in ${\rm erg\,s^{-1}\,Hz^{-1}\,cm^{-2}}$.

We divide the root-mean-squared deviation, as calculated above, by the average flux density of the rotator within the 1\,$\mu$m--5\,$\mu$m interval and minimize the resulting quantity over all nonrotating models.  This provides a dimensionless measure of the observable impact of rotation -- the smallest difference between a non-rotator's spectrum and a linear transformation of the rotator's spectrum.  This minimum normalized RMSD is a function of both rotation rate and orientation.

If there is a nonrotating model that exactly matches the spectral shape and features of a rotating model, then minimum ${\rm RMSD}$ for that model is zero.  If there is no nonrotating model that can match the spectral shape of the rotating model, then the minimum RMSD can be significantly nonzero.  

\subsection{Flux Anisotropy Factor} \label{anisotropy}

Rotation breaks the spherical symmetry of a brown dwarf, making the object's bolometric flux depend on the direction from which it is seen. As a result, if one assumes that a rotating dwarf is not rotating and thus spherically symmetric, one tends to over-estimate its luminosity from its flux, if one sees the object pole-on. On the other hand, an equator-on view leads to an under-estimate of the luminosity. In this section, we quantify this effect for a range of rotational speeds and rotational axis inclinations. 

Consider brown dwarf A with luminosity $L$ that we wish to know and a spectrum that we observe, with a certain bolometric flux. We allow for the possibility of rotation, anywhere between zero and critical. The true luminosity $L$ will not equal $4 \pi r^2$ times the bolometric flux (where $r$ is the distance) because the observed flux varies with viewing angle. Standard practice, however, is to infer the luminosity of a brown dwarf in exactly this way, by assuming isotropic radiation.  The flux density is measured at some wavelengths and extrapolated to wavelengths that are not measured and then integrated using a model of a nonrotating brown dwarf.

We reproduce a realistic luminosity inference by assuming that we have a nonrotating model B that matches the observed spectrum of brown dwarf A up to a scaling factor.  Model B has luminosity $L_{0}$, integrated flux ${\cal F}_{0}$, effective temperature $T_{0}$, radius $R_0$ and distance $r$. Let $p$ be the ratio of ${\cal F}_{0}$ and the model's total bolometric flux, i.e., the fraction of the bolometric flux within the wavelength range covered by the model spectrum. Given these definitions, the model's luminosity can be written as
\begin{equation} \label{eq:L0}
    L_{0} = 4 \pi \sigma_{\rm SB} R_0^2 T_0^4, 
\end{equation}
and its integrated flux -- as
\begin{equation} \label{eq:F0}
    {\cal F}_0 = \frac{1}{r^2}\, p\, \sigma_{\rm SB} R_0^2 T_0^4.
\end{equation}
We are interested in the ratio of the dwarf's true luminosity $L$ and the nonrotating model's luminosity $L_0$. Specifically, we wish to know what causes this ratio to deviate from one. We can write $L / L_0$ as
\begin{multline} \label{eq:LL0}
    \frac{L}{L_0} = \frac{L}{L_0} \frac{{\cal F}_0}{{\cal F}} \frac{{\cal F}}{{\cal F}_0} = \frac{L}{4 \pi \sigma_{\rm SB} R_0^2 T_0^4} \frac{\frac{1}{r^2}\,p\, \sigma_{\rm SB} R_0^2 T_0^4}{\cal F} \frac{{\cal F}}{{\cal F}_0} =\\ \frac{L\,p}{4 \pi r^2 {\cal F}} \times \frac{{\cal F}}{{\cal F}_0},
\end{multline}
where we have made two substitutions according to Equations \eqref{eq:L0} and \eqref{eq:F0}. Let us assume that $p$ is the same for the dwarf and the model. In other words, the latter are both at the same distance and require the same bolometric correction for flux that falls outside the observed wavelength range. The last expression of Equation \eqref{eq:LL0} consists of two multiplicative terms. The first term is the ratio of $L / (4\pi r^2)$, the bolometric flux we expect if the dwarf is nonrotating and spherically symmetric, to ${\cal F}/p$, an approximation of the dwarf's bolometric flux from the observed integrated flux and the fraction of the flux that is in the available wavelength range. The second term is the ratio of the dwarf's integrated flux to the model's integrated flux. 

If A does not rotate, then it has a well-defined radius $R$ and effective temperature $T$. In this case, its luminosity is
\begin{equation} \label{eq:L}
    L = 4 \pi \sigma_{\rm SB} R^2 T^4
\end{equation}
and its integrated flux is
\begin{equation} \label{eq:F}
    {\cal F} = \frac{1}{r^2}\, p\, \sigma_{\rm SB} R^2 T^4.
\end{equation}
Substituting for the first occurrences of $L$ and ${\cal F}$ on the right hand side of Equation \eqref{eq:LL0} according to Equations \eqref{eq:L} and \eqref{eq:F}, we obtain $L / L_0 = {\cal F} / {\cal F}_0$. In other words, if the dwarf does not rotate, then it is sufficient to multiply $L_0$ by the ratio of integrated fluxes ${\cal F} / {\cal F}_0$ to obtain $L$. 

On the other hand, if the dwarf rotates with non-zero velocity, multiplication of $L_0$ by the flux ratio ${\cal F}/{\cal F}_0$ does not, in general, give us $L$. We need to additionally multiply our estimate by $L\,p / (4 \pi r^2 {\cal F})$, which we call the anisotropy ratio, since it deviates from one due to the anisotropy of flux because of rotation-induced centrifugal deformation and gravity darkening. 

It turns out that the anisotropy ratio is mainly a function of the rotational speed and inclination and that it is otherwise almost independent of other brown dwarf parameters (like mass and temperature). To see this, consider Equation (31) in \citet{EspinosaLara_2011A&A}, which tells us that the surface flux at every location on the deformed dwarf's surface is proportional to $L / R_{\rm e}^2$, where $R_{\rm e}$ is the object's equatorial radius. If, additionally, the spatial flux distribution at each location is independent of temperature -- the way it is for a black body, this equation implies that multiplication of $L / R_{\rm e}^2$ by some factor multiplies the observed flux by the same factor and that the anisotropy ratio $L\,p / (4 \pi r^2 {\cal F})$ doesn't change with luminosity $L$, object size $R_{\rm e}$, or the effective temperature scale, which is proportional to $L / R_{\rm e}^2$. The object's mass $M$ does not enter Equation (31) in \citet{EspinosaLara_2011A&A} at all. Thus, we expect that the anisotropy ratio is almost entirely a function of rotational speed $\omega_{\rm R}$ and inclination $i$. The anisotropy ratio changes little with mass $M$ and, since the local dependence of intensity on viewing angle broadly matches that of a black body, we do not expect the anisotropy ratio to change significantly with luminosity $L$ or object size $R_{\rm e}$, either. In Section \ref{lum_cor} we will demonstrate this fact quantitatively for our models.

\section{Results} \label{results}

In this section, we present our findings -- the effects of rotation on brown dwarf spectra and the implications of these effects for the inference of brown dwarf parameters from nonrotating models.

\subsection{Spectral Intensity and Inferred Temperature}

The observed flux density of a rotating brown dwarf significantly depends on the inclination of its rotational axis. The temperature one infers for such an object from non-rotating models greatly depends on the inclination as well. In this section, we quantify these effects for a model inspired by one of the fastest-rotating known dwarfs.

\begin{figure*} 
\centering
\includegraphics[width=0.75\linewidth]{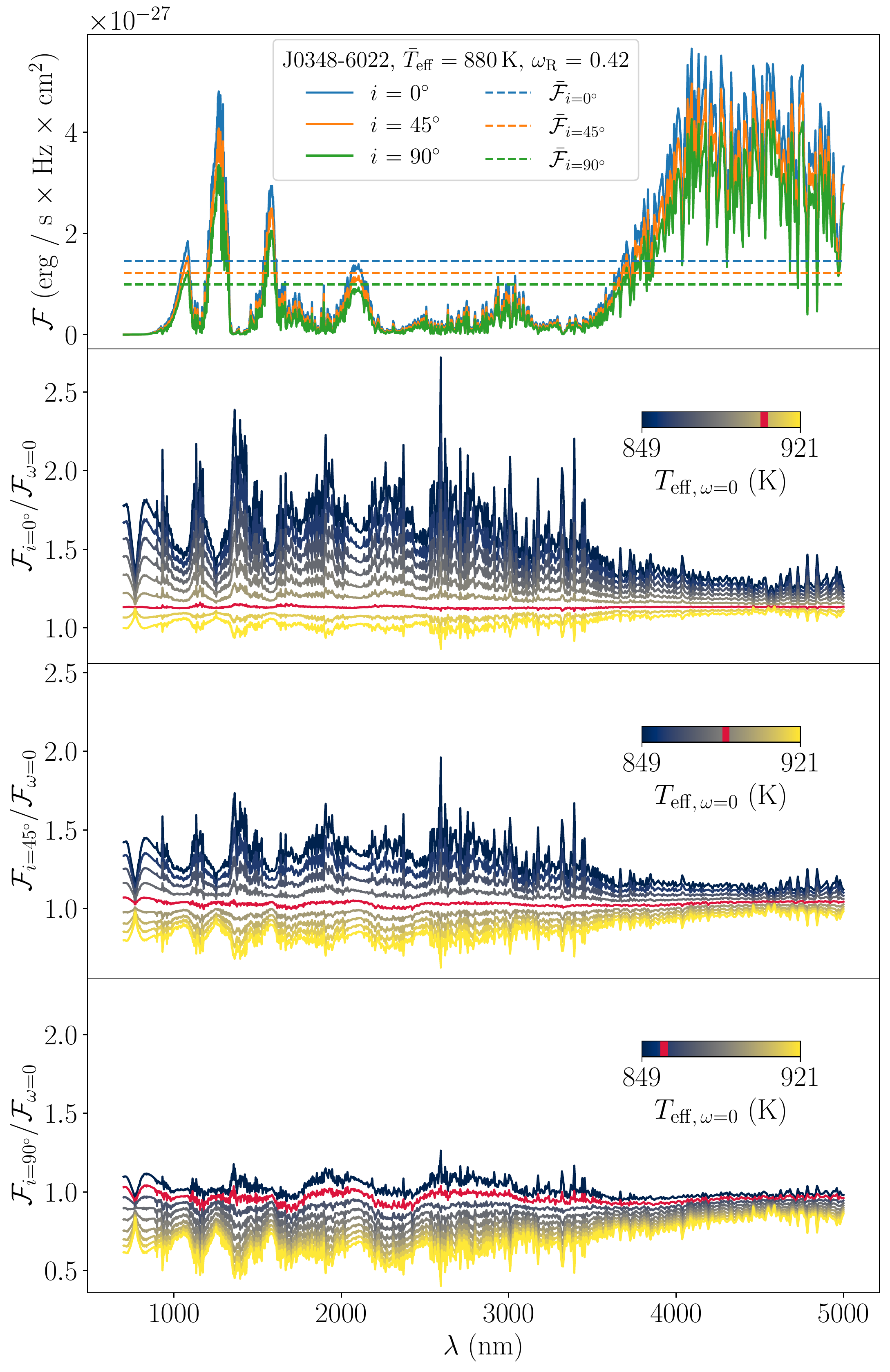}
\caption{Top panel: synthetic spectra of a fast-rotating brown dwarf model. Lower panels: ratios of the rotating model's pole-on spectrum to spectra of nonrotating models with effective temperatures between the equatorial (top line) and the polar (bottom line) temperatures of the rotator. Crimson marks the nonrotating dwarf with the smallest relative deviation of spectrum shape from the spectrum shape of the rotator. The range of the corresponding best-fit spectrum ratio -- indicating the detectability of rotation -- increases with inclination. }
\vspace{20pt}
\label{fig:t7}
\end{figure*}

\begin{figure*} 
\centering
\includegraphics[width=0.75\linewidth]{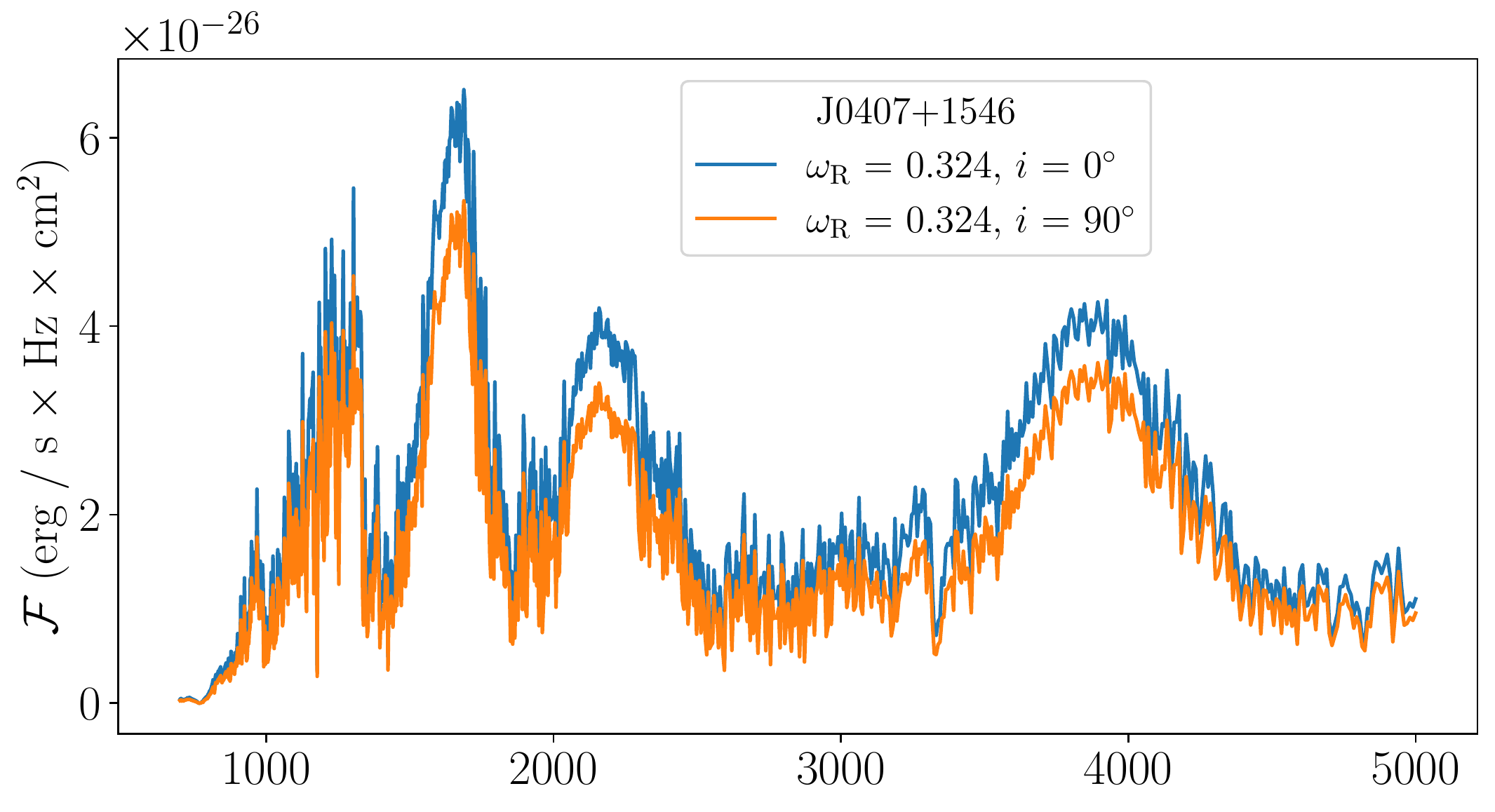}
\includegraphics[width=0.75\linewidth]{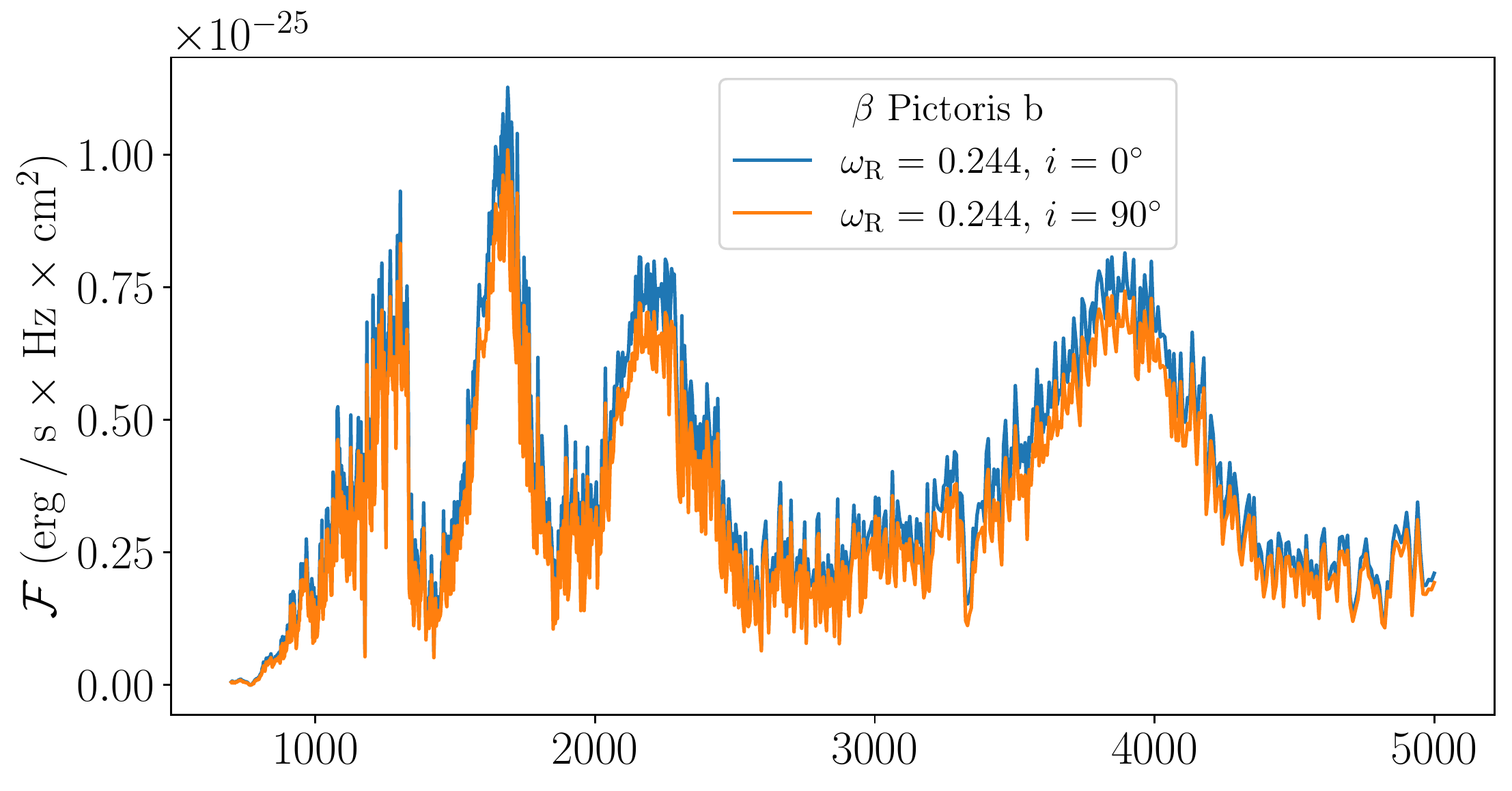}
\caption{Upper panel: synthetic spectra of the L3.5-type fast-rotating brown dwarf from \citetalias{Tannock_2021AJ}. Lower panel: synthetic spectra of $\beta$ Pictoris b. Both panels present spectra at the two extreme inclinations.}
\vspace{20pt}
\label{fig:extras}
\end{figure*}

Let us suppose that one of the Model 880 spectra in the upper panel of Figure \ref{fig:t7} is an observed spectrum of a rotating dwarf. There are portions of this spectrum with appreciable intensity where the flux differs by as much as a factor of 1.5 between zero degree inclination (pole-on view, maximum flux) and ninety degree inclination (equator-on view, minimum flux). Figure \ref{fig:extras} shows that the variation of spectral flux with inclination is less extreme for objects with lower rotational speed.

Next, we compare the synthetic observed rotator (Model 880) spectrum at every inclination to the library of associated nonrotating models (Section \ref{variants}). We search the library for the best-matching spectrum by minimizing the root-mean-square deviation (RMSD) between the nonrotating model spectrum and a linear transformation of the observed spectrum that brings it closest to the non-rotator (Section \ref{comparisons}). 

It is apparent from Figure \ref{fig:t7} that such a procedure leads to cooler inferred temperatures when the view of the rotator is closer to equator-on and hotter temperatures when it is towards pole-on. The inferred temperature is always in the range between the maximum, polar temperature of the dwarf and its minimum, equatorial temperature. 

\subsection{Rotation Detectability}

\begin{figure*} 
\centering
\includegraphics[width=0.75\linewidth]{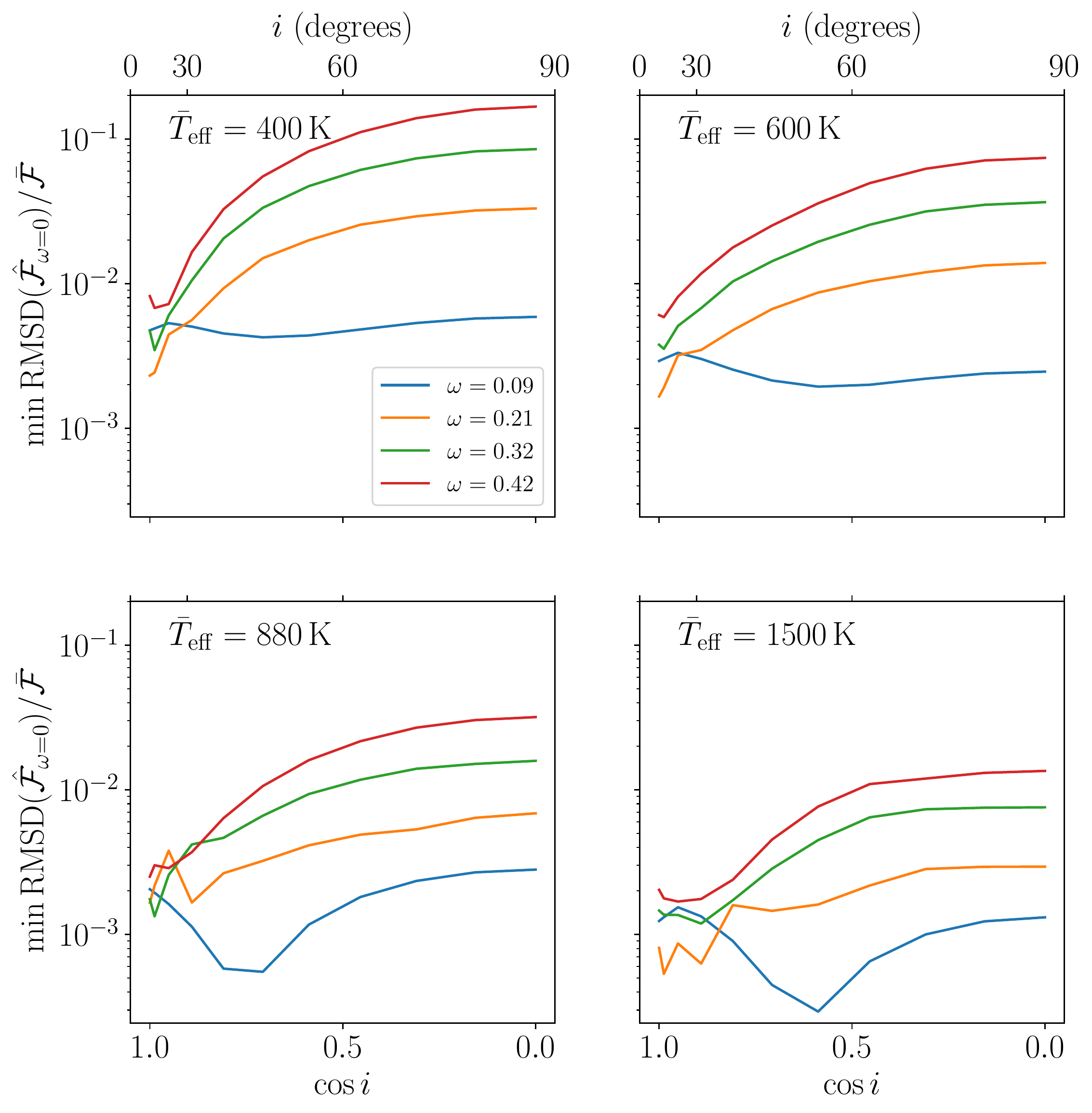}
\caption{Normalized minimum root mean squared difference (RMSD) between a nonrotating dwarf's spectrum and the closest linear transformation of a rotator's spectrum, minimized across nonrotating models for a given rotator. This is a measure of the difference between the rotator's spectrum shape and the spectral shape of the best-matching non-rotator. The horizontal axis is the cosine of the rotational axis inclination.  The different panels correspond to different average surface temperatures of the rotating dwarf. Comparison of the panels indicates that lower temperatures correspond to higher relative differences between the rotating and the nonrotating dwarf spectra, and thus, potentially, easier detectability of the rotational effects. Additionally, the panels suggest easier detectability at higher inclinations, in agreement with Figure \ref{fig:t7}.}
\vspace{20pt}
\label{fig:rmsd}
\end{figure*}

In addition to spectral intensity and inferred temperature, inclination affects the detectability of rotation from spectral shape. The minimum RMSD from the previous section can serve as a quantitative measure of detectability, where higher minimum RMSD corresponds to higher detectability of rotation. At pole-on inclination, there is a nonrotating model with a spectrum that is very similar to the rotator's spectrum, both according to the minimum RMSD criterion (Figure \ref{fig:rmsd}) and according qualitative examination (see Figure \ref{fig:t7}). At the same time, Figures \ref{fig:t7} and \ref{fig:rmsd} show that the more equator-on inclinations produce spectra that are increasingly distinguishable from all non-rotator spectra. 

Figure \ref{fig:rmsd} shows that the detectability of rotation from spectral shape grows not only with increasing inclination, but also with decreasing temperature. Specifically, this figure demonstrates that the minimum RMSD measure, normalized by the average integrated flux, decreases as one proceeds from cooler to hotter models. This may be related to the fact that cooler models show stronger departures from blackbody spectra, with corresponding increases in the sharpness of temperature-dependent spectral peaks and troughs.

The horizontal axis of Figure \ref{fig:rmsd} is linear in the cosine of inclination, $\cos{i}$. This way, equal-space horizontal axis intervals correspond to equal probabilities of brown dwarfs under a spatially isotropic distribution of rotational axis directions, which corresponds to observed inclination probability density that is proportional to $\sin{i}$. 

Figure \ref{fig:rmsd} shows that the normalized RMSD reaches 2\% for Model 880, suggesting that, if spectral precision and modeling fidelity reach this level, rotation may be detectable from spectra alone. The detectability is greatest for edge-on orientations, so that a substantial $v_{\rm e}\sin{i}$ would further establish the spectrum as that of a rapidly rotating brown dwarf. 

\subsection{An Empirical Luminosity Correction} \label{lum_cor}

In this section we quantify the dependence of the inferred luminosity on inclination and rotation rate.  To do this, we first create nine rotating models associated with Model 880, as described in Section \ref{variants}. Their rotation rates $\omega_{\rm R}$ are equally spaced between 0.1 and 0.5 and their luminosities are all equal to that of Model 880. We calculate spectra at several inclinations for each model, and in each case produce the integrated flux ${\cal F}$. For every such spectrum, we find the nonrotating Model 880 variant with the maximally similar spectrum, according to the RMSD criterion. Let us say that this nonrotating model has integrated flux ${\cal F}_0$ and luminosity $L_0$. We then compute a product of luminosity and flux ratios on the left hand side of the following equation, which is a re-arrangement of Equation \eqref{eq:LL0}:
\begin{equation} \label{eq:LL0_F0F}
    \frac{L}{L_0} \times \frac{{\cal F}_0}{{\cal F}} = \frac{L\,p}{4 \pi r^2 {\cal F}}.
\end{equation}
\begin{figure} 
\centering
\includegraphics[width=\linewidth]{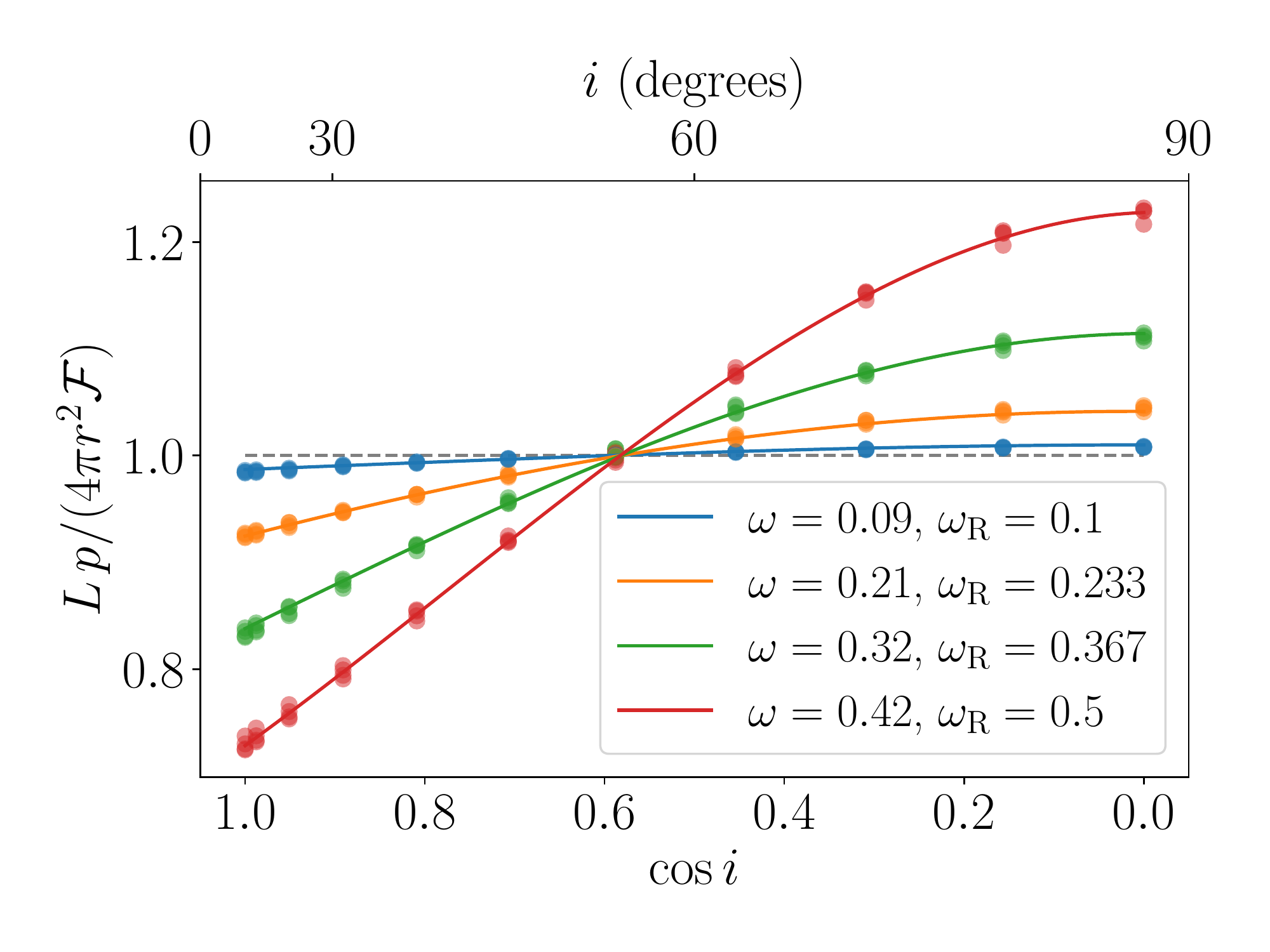}\\
\caption{Luminosity correction factor due to the rotational anisotropy of bolometric flux. This factor is a ratio. The numerator of the ratio is $L / (4 \pi r^2)$, the flux we would expect from the the dwarf's luminosity $L$ and distance $r$ if the flux were isotropic -- the way it is for a nonrotating, spherically symmetric object. The denominator of the ratio is the observed bolometric flux ${\cal F} / p$ of the rotating dwarf. Here, ${\cal F}$ is the dwaf's flux between the wavelengths of 0.7 and 5.0 ${\rm \mu m}$, whereas $p$ is the proportion of a non-rotator's flux that is between these wavelengths. When $i = 55^\circ$, the effect is non-existent. At this inclination, the bolometric flux is equal to the direction-averaged flux, no matter what the rotational speed is. The discrete markers are due to our synthetic spectra at $\bar{T}_{\rm eff} \in \{400, 600, 880, 1500\}\, {\rm K}$. The lines are cubic fits in $\cos{i}$ for $\bar{T}_{\rm eff} = 880\,{\rm K}$, with coefficients' quadratic dependence on $\omega_{\rm R}$ given in Section \ref{lum_cor}. The effect mainly depends on the rotational speed $\omega_{\rm R}$ and inclination $i$.}
\label{fig:rat_lum}
\end{figure}

Figure \ref{fig:rat_lum} presents the anisotropy ratio, i.e. the expression on either side of Equation \eqref{eq:LL0_F0F} for four of the rotational speeds and all the inclinations that we use. 

The lines in Figure \ref{fig:rat_lum} are cubic polynomials in the cosine of inclination $u = \cos{i}$, whose coefficients are given by quadratic fits to all nine rotational speeds $\omega_{\rm R}$. The corresponding full coefficient matrix is
\begin{equation}
    \mathbf{A}
    =
    \begin{bmatrix}
    3.0791 & -0.9858 & 0.0807 \\
    -4.5216 & 0.7383 & -0.0644 \\
    -0.3774 & 0.1489 & -0.0113 \\
    1.1536 & -0.1482 & 1.0133
    \end{bmatrix}.
\end{equation}
Thus, the anisotropy ratio in Figure \ref{fig:rat_lum} at $\omega_{\rm R}$ and $u = \cos{i}$ is given by 
\begin{equation}
    \begin{bmatrix}
        u^3 & u^2 & u & 1
    \end{bmatrix}
    \mathbf{A} 
    \begin{bmatrix}
        \omega_{\rm R}^2 \\ \omega_{\rm R} \\ 1
    \end{bmatrix}.
\end{equation}
The maximum difference between the fits and the discrete anisotropy ratios in Figure \ref{fig:rat_lum} is 0.01, i.e., the fitting formula can correct the luminosity with a maximum error of 1\%. This maximal difference is found at $\omega_{\rm R} = 0.5$, $i = 90^\circ$, and $\bar{T}_{\rm eff} = 1500\,{\rm K}$: a hot, very rapid rotator seen equator-on. The root-mean-square difference between the fit lines and all the discrete points in this figure is $1.2\times10^{-5}$, much less than 1\%. The fact that these differences are small in comparison with the range of the anisotropy ratio confirms the proposition in Section \ref{anisotropy} that the ratio is largely independent of object temperature and that it is mostly a function of $\omega_{\rm R}$ and $i$.

Section \ref{anisotropy} interprets the anisotropy ratio as a correction factor one must use in addition to the ratio of integrated fluxes in order to obtain a rotator's true luminosity from the luminosity of a best-matching model. Figure \ref{fig:rat_lum} shows that this correction factor changes the luminosity by as much as 20\% at extreme inclinations and rotational velocities that are typical of fastest-rotating dwarfs.

\section{Discussion and Conclusions} \label{discussion}

Rotational period and projected equatorial velocity measurements for substellar objects frequently translate to significant fractions of critical rotation rates \citep{Tannock_2021AJ,Chilcote_2017AJ,Snellen_2014Nature,Helled_2009P&SS,Eisloffel_2007IAUS,Crossfield_2014A&A}. The centrifugal expansion of a rotating object's equatorial regions will produce a significant variation in its surface temperature. As a result, we expect both the shape of a rotator's observed spectrum and its total observed flux to depend on the object's orientation. Furthermore, we expect this effect to increase with greater rotational speed, since the latter leads to a greater temperature contrast between the equator and the poles.

In this work, we explore the dependence of a substellar object's observables on its rotational speed and rotational axis inclination. We also place this dependence in the context of a comparison with the expected observables of nonrotating models. In order to accomplish these tasks, we use \texttt{PICASO} (a Planetary Intensity Code for Atmospheric Spectroscopy Observations) to generate synthetic spectra and process them with \texttt{PARS} (Paint the Atmospheres of Rotating Stars), which produces spectra of rotating, self-gravitating, gaseous masses at different rotational speeds and inclinations \citepalias{Lipatov_2020ApJ}. 

An initial analysis shows that the specific flux of a typical fast-rotating brown dwarf can differ by as much as a factor of 1.5 between the two extreme inclinations in spectral regions with appreciable intensity (see the upper panel of Figure \ref{fig:t7}, for example). 

Next, we ask whether the shape of a rotationally deformed substellar object's spectrum differs significantly from the spectra of similar objects that are nonrotating and therefore spherically symmetric. In other words, we wish to know whether one can infer rotation from the shape of an object's spectrum alone. To this end, we use \texttt{PARS} to calculate synthetic spectra of a typical quickly rotating brown dwarf at different rotational axis inclinations, as well as spectra of nonrotating objects that are otherwise similar to the rotator. We find that, when we observe the rotator pole-on, its spectrum is virtually indistinguishable from that of another, nonrotating object at a certain temperature between the rotator's equatorial and polar values. On the other hand, as the rotator approaches an equator-on view, its spectrum becomes increasingly different from that of any nonrotating object. We demonstrate this effect in Figure \ref{fig:t7}.

To quantify the difference between the spectral shapes of a rotator and a non-rotator, we compute the root-mean-square difference between the non-rotator's specific flux at different wavelengths and the closest linear transformation of the rotator's flux. We minimize this quantity across non-rotators and divide it by the rotator's mean flux, to obtain a dimensionless measure of rotation's detectability from spectral shape. We compute this minimized RMSD measure across different average rotator temperatures, rotational velocities, and inclinations. Figure \ref{fig:rmsd} shows that, in accordance with the qualitative results of Figure \ref{fig:t7}, detectability of rotation from spectral shape increases when the rotational speed is greater and the view is closer to equator-on. In addition, Figure \ref{fig:rmsd} indicates that detectability increases when the temperature of the rotating dwarf decreases.

A nonrotating model is spherically symmetric. Thus, its observed flux does not depend on the spatial direction to the observer. On the other hand, a rotator's flux is anisotropic -- it depends on the observer's direction or, equivalently, on the inclination of the rotational axis with respect to the observer's view. Thus, although the luminosity of a nonrotating object can be inferred from its observed flux, a rotator's luminosity estimate requires an additional correction due to the anisotropy effect. Without this correction, the luminosity of an equator-on object is under-estimated, while that of a pole-on object is over-estimated. We calculate this correction for a variety of rotating models and plot it in Figure \ref{fig:rat_lum}. We find that it is relatively insensitive to average object temperature. On the other hand, the correction depends strongly on rotational speed and inclination. It reaches its extreme values at the two extreme inclinations, where it adjusts luminosity estimates by as much as $\sim$20\% near the rotational speeds of fastest-rotating observed dwarfs. Section \ref{lum_cor} provides an approximation of the luminosity correction as a function of rotational speed and inclination.

\section*{Data Availability}

The brown dwarf atmosphere intensity grid in this article, produced by the \texttt{PICASO} radiative transfer code and based on Sonora model atmospheres, is available on Zenodo, at https://doi.org/10.5281/zenodo.6842801. The version of \texttt{PARS} that took this intensity grid as input and produced all the figures in the article is also on Zenodo, at https://doi.org/10.5281/zenodo.6842745.\\

\noindent {\it Software}: {PARS \citep{Lipatov_2020ApJ}, The NumPy Array \citep{vanDerWalt_2011}, Matplotlib \citep{Hunter_2007}, PICASO \citep{picaso2022}.\\}

\bibliography{refs.bib}
\bibliographystyle{mnras}
\label{lastpage}
\end{document}